\documentclass[aip,jcp,amsmath,amssymb,reprint,floatfix,notitlepage,longbibliography]{revtex4-1}

\usepackage{graphicx}
\usepackage{dcolumn}
\usepackage{bm}
\usepackage{placeins} 
\usepackage{tabularx}
\usepackage[version=3]{mhchem} 
\usepackage{array}
\usepackage{natmove}
\usepackage{mciteplus}
\usepackage{graphicx}
\usepackage{xr}
\usepackage{xcolor}
\usepackage[normalem]{ ulem }
\usepackage{soul}
\usepackage{hyperref}
\usepackage{longtable}
\hypersetup{%
  colorlinks=true, 
  breaklinks=true, 
  urlcolor=blue, 
  linkcolor=black, 
  citecolor=black, 
}
\makeatletter
\providecommand*{\input@path}{}
\g@addto@macro\input@path{{./Figures/}}
\makeatother

\graphicspath{{.}{Figures/}}

\begin{document}

\title{Noble gases in carbonate melts: constraints on the solubility and the surface tension by molecular dynamics simulation}

\author{Elsa Desmaele}\email{elsa.desmaele@gmail.com}
\affiliation{Sorbonne Universit\'e, CNRS, Laboratoire de Physique Th\'eorique de la Mati\`{e}re Condens\'ee, LPTMC, F75005, Paris, France}
\author{Nicolas Sator}
\affiliation{Sorbonne Universit\'e, CNRS, Laboratoire de Physique Th\'eorique de la Mati\`{e}re Condens\'ee, LPTMC, F75005, Paris, France}
\author{Bertrand Guillot}\email{guillot@lptmc.jussieu.fr}
\affiliation{Sorbonne Universit\'e, CNRS, Laboratoire de Physique Th\'eorique de la Mati\`{e}re Condens\'ee, LPTMC, F75005, Paris, France}

\begin{abstract}
Although they are rare elements in the Earth's mantle, noble gases (NG) owe to their strongly varying masses (a factor $>50$ from He to Rn) contrasting physical behaviors making them important geochemical tracers. When partial melting occurs at depth, the partitioning of NGs between phases is controlled by a distribution coefficient that can be determined from the solubility of the NGs in each phase.\\
Here we report quantitative calculations of the solubility of He, Ne, Ar and Xe in carbonate melts based on molecular dynamics simulations. The NG solubilities are first calculated in \ce{K2CO3}-\ce{CaCO3} mixtures at 1 bar and favorably compared to the only experimental data available to date. Then we investigate the effect of pressure (up to 6~GPa), focusing on two melt compositions: a dolomitic one and a natrocarbonatitic one (modeling the lava emitted at Ol Doinyo Lengai). The solubility decreases with the amount of alkaline-earth cation in the melt and with the size of the noble gas. In the natrocarbonatitic melt, Henry's law is fulfilled at low pressures (up to $\sim 0.1$~GPa). At higher pressures (a few GPa) the solubility levels off or even starts to diminish smoothly (for He at $P > 2$ GPa and Ar at $P > 4$ GPa). In contrast, in molten dolomite the effect of pressure is negligible on the studied $P$ range ($3-6$~GPa).
At the pressures of the Earth's mantle, the solubilities of noble gases in carbonate melts are still of the same order of magnitude as the ones in molten silicates ($10^0-10^1$ mol\%). This suggests that carbonatitic melts at depth are not preferential carriers of noble gases, even if the dependence with the melt composition is not negligible and has to be evaluated on a case-by-case basis.\\
Finally we evaluate the surface tension at the interface between carbonate melts and noble gases and its evolution with pressure. Whatever the composition of the melt and of the NG phase, the surface tension increases (by a factor $\sim 2$) when $P$ increases from 0 to 6~GPa. This behavior contrasts with the situation occurring when \ce{H2O} is in contact with silicate melts (then surface tension drops when pressure increases to a few GPa).
\end{abstract}
\maketitle
\section{Introduction}
By the time of accretion, noble gases were already present in proto-Earth material.\cite{Marty2012} Their concentration in the mantle has since evolved through the competing effect of volcanic degassing and radioactive decay. Although they are inert species their strongly varying masses in the series from He to Rn confer them contrasting physical behaviors in the Earth's mantle. Hence these elements (and their isotopes even more so) constitute important geochemical tracers.\cite{Moreira2013} More specifically, noble gases are incompatible in mantle rocks and tend to partition into the liquid phase when partial melting occurs at depth. In the presence of two immiscible liquids, the partioning is determined by a distribution coefficient, that is related to the ratio of the solubility of the noble gas in each phase. For example a carbonatite-silicate  immiscibility is currently the most probable scenario for the genesis of the natrocarbonatites emitted by Ol Doinyo Lengai in Tanzania \cite{Fischer2009}. In a more general perspective, the noble gas systematics provides an insight into the past and present dynamics of the carbon-bearing phases (silicates and/or carbonates) in the Earth's mantle (\emph{e.g.} the carbon cycle). Hence solubility data under high pressure and in different magmatic liquids, are requisite in order to study these phenomena. Whereas many studies have been devoted to the solubility of noble gases in molten silicates (mainly at low pressures),\cite{White1989,Montana1993,Carroll1993,Chamorro1996,Schmidt2002,Bouhifd2008} very little is known concerning their behavior in carbonate melts which are yet of particular interest.\cite{Dasgupta2006,Dasgupta2010,Burnard2010,Hammouda2015} \\
To date, the only experimental constraints are given by Burnard \emph{et al.}\cite{Burnard2010}, who measured the solubility of noble gases in quenched carbonate melts.  In this approach the concentration in the liquid is assumed to be the same as in the glass resulting from the quenching. Moreover the quenching is considered to be fast enough to avoid crystal nucleation. The study reported the solubility of He and Ar in some \ce{K2CO3}-\ce{CaCO3} mixtures at 1 bar between 1123 and 1223~K. The authors also attempted to study Mg-bearing carbonate melts, but could not succeed in quenching them into a glass.\\
By contrast, in a molecular dynamics (MD) simulation the liquid phase can always be studied, regardless of its composition (and of the pressure/temperature condition). The relevance of the liquid properties calculated from a MD simulation relies on the quality of the implemented interaction potential or force field (FF). The adjustment of the FF generally is system-specific. In a previous study we have presented a FF to model carbonate melts in the \ce{CaCO3}--\ce{MgCO3}--\ce{K2CO3}--\ce{Na2CO3}--\ce{Li2CO3} system, and demonstrated that it leads to an accurate reproduction of their thermodynamics, structure and transport properties \cite{moi2018,moi2019}. Here we perform molecular dynamics simulations based on this FF to study the solubility of noble gases (from He to Xe) into \ce{K2CO3}-\ce{CaCO3} mixtures (for comparison with the data of Burnard \emph{et al.}\cite{Burnard2010}) and into dolomitic and natrocarbonatitic melts to investigate the evolution with pressure (up to 6 GPa). First we present in section~\ref{smethod} the method to evaluate the noble gas solubility. The results are discussed in section~\ref{ssol}. The surface tension of carbonate melts in contact with a noble gas fluid is evaluated and discussed in section~\ref{sstens}.

\section{Method}\label{smethod}
\begin{figure}[hbt]
\centering
\includegraphics{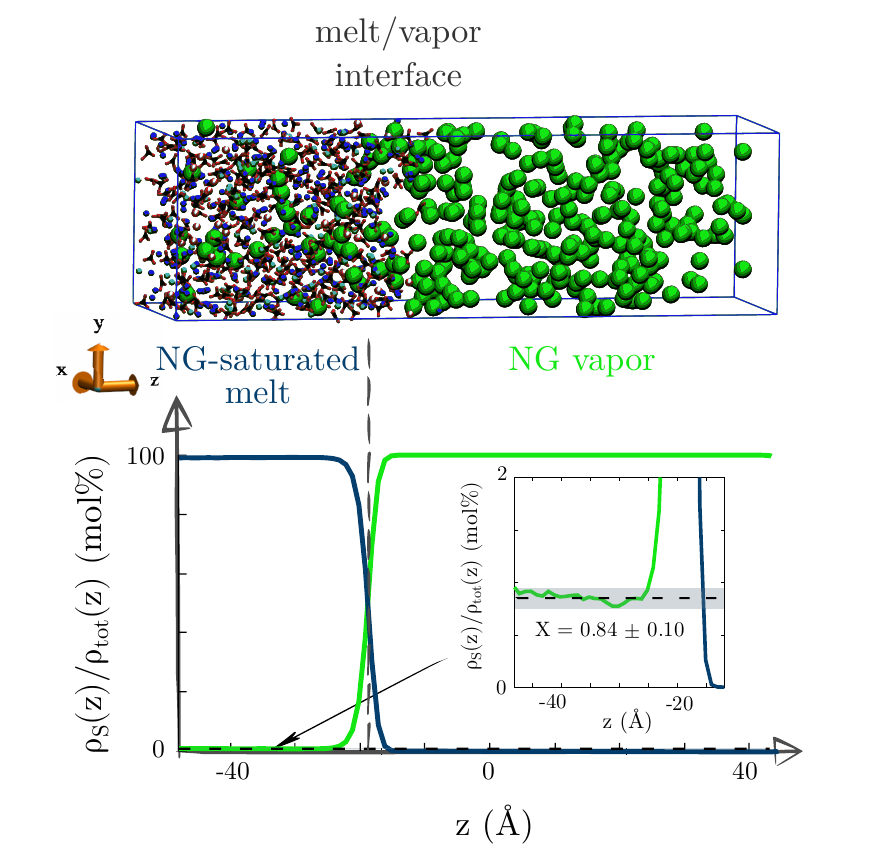}
\caption{Determination of the solubility of a noble gas (e.g. Ne in green) in  a carbonate melt (e.g. natrocarbonatite at 1600~K and 0.1~GPa) by simulating explicitly the liquid/gas interface (up). The density profile of the noble gas component is established perpendicularly to the interface (down). The solubility $\rho_s (z)/\rho_{tot} (z)$ of the species $s$ (Ne in green or melt in blue) is calculated from the plateau value of the density profile. The inset emphasizes the density profile of the noble gas in the melt.}
\label{fdprofile}
\end{figure}
\subsection{Force Field}
To describe the interactions within the carbonate melt we used the interatomic FF proposed by Desmaele \emph{et al.}\cite{moi2018,moi2019} The interactions between noble gas (NG) atoms were modeled by the FF used by Guillot and Sator\cite{Guillot2012}. These Buckingham potentials (see Eq.~(\ref{ebuck}) in Appendix~\ref{a-FF}) were adjusted on the accurate potential energy surface derived by Tang and Toennies\cite{Tang2003}. As for the interactions between the NG and the carbonate melt elements, they are described by two contributions, a NG-cation and a NG-carbonate one, both of them being modeled by Lennard-Jones (LJ) potentials (Eq.~(\ref{elj})). For the NG-cation interactions we used the LJ parameters determined by \citet{Guillot2012} for the \{silicate~+~NG\} system. For the  interaction between the NG and the oxygen atoms of the carbonate ions, we retained the parameters established by  Aubry \emph{et al.}\cite{Aubry2013} for the system \{silicate~+~CO$_2$~+~NG\} and set the interactions between the NG and the carbon atom of the carbonate anion to zero, as the carbon is considered to be screened by the surrounding oxygen atoms \cite{moi2018,moi2019} (in fact, the carbon atom is deeply embedded  into the oxygen electronic clouds). The choice of such a force field is based on the assumption that the interaction potentials between NGs and melts cited above are transferable from a melt to another (e.g. \ce{CO2}-bearing silicate to carbonate). It is justified a posteriori by the agreement between the calculated solubilities and the solubility data of Burnard \emph{et al.}\cite{Burnard2010} (see Section \ref{sec:okgr}). 

\subsection{Solubility calculations}
\subsubsection{Test Particle Method (TPM)}		
The solubility, expressed as a molar fraction, of a noble gas in a melt is given by (see Appendix~\ref{a-thermo}):
\begin{equation}
X=\cfrac{\cfrac{\rho_v}{\rho_m} \cdot \cfrac{\Gamma_m}{\Gamma_v}}{1+\cfrac{\rho_v}{\rho_m} \cdot \cfrac{\Gamma_m}{\Gamma_v}} \enspace  ,
\label{egammax}
\end{equation}
where $\rho_m$ and $\rho_v$ are the number densities of the pure melt and of the vapor phase (noble gas), respectively, and $\Gamma_m$ and $\Gamma_v$ are the solubility constants of the noble gas in the melt and in its pure phase, respectively.\\
At low pressure $P$, the gas phase can be considered as ideal, allowing to retrieve Henry's law from Eq.~(\ref{egammax}): 
\begin{equation}
X=P/k_H \enspace ,
\label{ehenry}
\end{equation}
where $k_H = \rho_m k_B T / \Gamma_m$  is the Henry constant. The solubility of noble gases in the melt was calculated by the Test Particle Method (TPM) introduced by \citeauthor{Widom1963}.\cite{Widom1963} This method allows to determine the excess chemical potential of a solute in a solvent, as a function of the potential energy distribution seen by the solute inserted in the solvent (at infinite dilution), namely
\begin{equation}
\mu^{ex, i} = -k_B T \ln \langle \mathrm{e}^{-\Psi_i /k_BT} \rangle_0  \enspace  ,
\end{equation}
where $\mu^{ex, i}$ is the excess chemical potential of a solute $i$ (e.g. a noble gas), $k_B$ is the Boltzmann constant, $T$ is the temperature, $\Psi_i$ is the interaction energy between the solute $i$ and the solvent (e.g. a carbonate melt) in which it is solvated, and $\langle \cdots \rangle_0$ means that a statistical averaging is done on the solvent molecules only, at a given ($T,P$) condition. Notice that the solute particle acts as a ghost particle, and that many insertions of the latter particle in the solvent configurations are needed to accurately evaluate $\mu^{ex}$ (for further details see \citet{Guillot2012} and Appendix~\ref{a-TPM}).\\
In practice a set of microscopic configurations of the melt is first generated by MD. Then the insertion of the solute particle (the noble gas atom) is attempted many times into the MD-generated configuration. For each insertion of the solute particle the interaction energy $\Psi$ is calculated.\\
The solubility parameter $\Gamma$ is then evaluated by the averaging over all the attempted insertions 
\begin{equation}
\begin{split}
\Gamma & = \mathrm{e}^{-\mu^{ex}/k_BT}= \langle \text{e}^{-\Psi/k_BT}\rangle \\& =\frac{1}{N_\text{test}}\sum_i^{N_\text{test}}\text{e}^{-\Psi_i/k_BT} \enspace , \end{split}
\label{egammanvt} 
\end{equation}
where $N_\text{test}$ is the number of attempted insertions. According to this method, the solubility parameter $\Gamma$ can be evaluated in the two coexisting phases: the carbonate melt and the noble gas phase (providing $\Gamma_m$ and $\Gamma_v$), and the solubility of the NG can be easily obtained from Eq.~(\ref{egammax}). 
\subsection{Explicit Interface Method (EIM)}
As an alternative to the TPM, the solubility of a gas in a melt can be calculated from a numerical experiment which consists in simulating explicitly the equilibrium between the two phases (for further details see~\citet{Guillot2011,Guillot2012}). The solubility is then simply obtained by counting the average number of noble gas atoms in the melt, once equilibrium is reached between the two phases, i.e. when the liquid is saturated in gas (Figure~\ref{fdprofile}).

\subsection{Complementarity between the two methods}
\label{sec:eqsol}
Depending on the thermodynamic conditions of interest, one of the two methods presented above is the most appropriate one to calculate the solubility. Thus in the framework of the TPM, the solubility parameter $\Gamma_m$ is theoretically defined at infinite dilution, which means that it is only meaningful for low concentrations of NG in the melt. On the contrary the explicit interface method (EIM) becomes useful when the pressure of the gas phase is sufficiently high for the simulated melt to accommodate a significant number of NG atoms. In brief, the TPM will be used at low pressures and the EIM at high pressures.\\  
In this context we have checked that the two methods give results that are consistent with one another. For example in natrocarbonatite (Na$_{1.1}$K$_{0.18}$Ca$_{0.36}$CO$_3$) at 1600~K and 0.1~GPa: $X^{TPM}_\text{Ne}= 0.78 \pm 0.01 $~mol\% (using Eq.~(\ref{egammax}) with $\Gamma_m=16.66 \pm 0.04 \times 10^{-3}$, $\Gamma_v=820.5 \pm 0.1 \times 10^{-3}$, $\rho_m = 17.98$~mol/L and $\rho_v = 6.947$~mol/L) and  $X^{EIM}_\text{Ne}= 0.84 \pm 0.10$~mol\%. Note that the rather large uncertainty on $X^{EIM}_\text{Ne}$ (see Figure~\ref{fdprofile}) is due to the relatively low pressure (0.1~GPa) considered here. 

\section{Solubility}\label{ssol}
\label{sec:okgr}
\subsection{Low pressure}	
\begin{figure}[ht]
\centering
\includegraphics{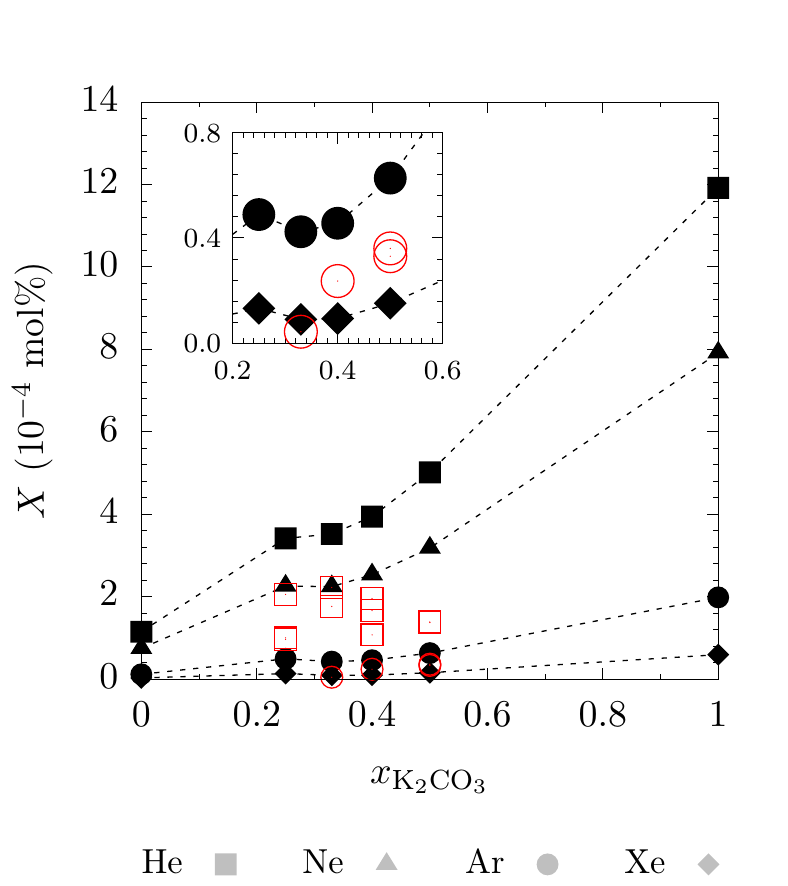}
\caption{Molar fraction of He, Ne, Ar and Xe solubilized in various \ce{K2CO3}--\ce{CaCO3} mixtures at 1173~K and 1~bar. The TPM results from this study are represented by plain black symbols, the measurements of Burnard \emph{et al.}\cite{Burnard2010} by red empty symbols. The error bars for the MD results are of the order of the size of the symbols (see text). The inset focuses on the region of low values for Ar and Xe solubilities.}
\label{fkcgr}
\end{figure}

First we focus on the mixtures studied by Burnard \emph{et al.}\cite{Burnard2010}, namely  \ce{K2CO3}--\ce{Na2CO3} mixtures at 1173~K and 1~bar. The solubilities calculated by  MD by the means of the TPM, and the ones measured by Burnard \emph{et al.}\cite{Burnard2010} are reported on Figure~\ref{fkcgr}. For some mixtures Burnard \emph{et al.}\cite{Burnard2010} have made several measurements of the solubility of helium. The error bars on each measurement is low and comprised within the symbol of Figure~\ref{fkcgr}. However the data dispersion for the He solubility at a given melt composition suggests that the uncertainty on the experimental results is actually greater than 50~\%. According to the authors, this could be due to the fast diffusion of the He gas out of the glass upon quenching. This He loss is likely systematic although its magnitude is varying. So the solubility data given by Burnard \emph{et al.}\cite{Burnard2010} for He should be considered as a lower limit. Hence the results of our simulations for He are compatible with the measurements of Burnard \emph{et al.}\cite{Burnard2010}. In contrast, for Ar the agreement between the TPM results and the experimental data is better. This is consistent with the assumption made for He (gas loss), because Ar being a heavier element it less easily escapes out of the quenched sample. \\
Beside He and Ar, we report the solubility for Ne and Xe. For a given composition of the liquid, the solubility increases as the atomic radius of the NG decreases. This behavior is similar to the one observed in silicates, where the solubility is related to the ionic porosity  \cite{Guillot2012}. Such a solvation mechanism can be termed as entropic, as it is mainly related to the structural disorder and cavity formation in the liquid. In contrast, in polar solvents (e.g. water) the solvation of NG is enthalpically driven with a solubility that increases with the polarizability of the solute \cite{Smith1983,Guillot1993} (i.e. with the size of the NG).\\
In their study, Burnard \emph{et al.}\cite{Burnard2010} reported that the solubility is hardly sensitive to the composition of the melt.  We believe that the high uncertainties on their data as well as the narrowness of the studied composition range ($0.2<x_{\ce{K2CO3}}\leq0.5$) are accountable for this observation. On the contrary MD simulations reveal that the composition of the melt has a strong effect (especially for $ x_{\ce{K2CO3}} < 0.25$ and $ x_{\ce{K2CO3}} > 0.4$), regardless of the noble gas species that is considered. In fact the solubility decreases continuously  when the content of calcite increases. For instance the solubility of He contrasts by an order of magnitude between the two end-members \ce{K2CO3} end \ce{CaCO3} (see Figure \ref{fkcgr}).
\subsection{Evolution with pressure}
\begin{figure}
\centering
\includegraphics{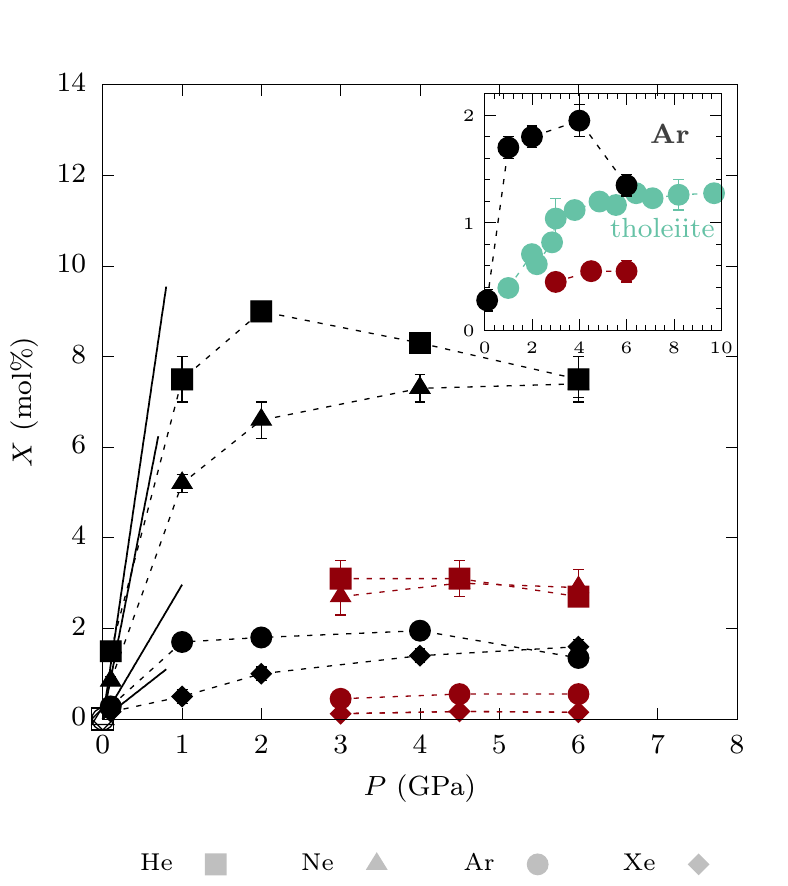}
\caption{Solubity of He, Ne, Ar and Xe (in molar fraction) in natrocarbonatite at 1600~K (black) and in dolomite at 1673~K (brown) as a function of pressure. The dotted lines are guides to the eye. Empty symbols represent the results of the TPM at 1~bar  for He (0.0012~mol\%), Ne (0.00089~mol\%), Ar (0.00030~mol\%) and Xe (0.00014~mol\%) in natrocarbonatite. The black lines represent Henry's law deduced from the latter values. The inset compares the solubility of Ar calculated in natrocarbonatite (black) and in dolomite (brown) to the one measured in tholeiite (green) by \citeauthor{Schmidt2002}.\cite{Schmidt2002}}
\label{fsol-gr} \end{figure} 
Simulating a biphasic system (EI method) enables to calculate the NG solubility under high pressures. We focus on two carbonate melts: molten dolomite (\ce{Ca_{0.5}Mg_{0.5}CO3}) at 1673~K and a \ce{Na2CO3}--\ce{K2CO3}--\ce{CaCO3} mixture (in proportions 55, 9 and 36 mol\%, respectively) at 1600 K, modeling the natrocarbonatite emitted at Ol Doinyo Lengai.\cite{Keller2012,moi2019} Figure \ref{fsol-gr} reports the solubility of He, Ne, Ar and Xe calculated at pressures ranging from 3 to 6~GPa for dolomite, and from 0.1 to 6~GPa for the natrocarbonatite. The solubilities in the natrocarbonatite were also calculated at 1~bar using the TPM. \\
Over the whole studied pressure range, the solubility of a given NG is greater in the natrocarbonatite than in molten dolomite. This is consistent with the decrease of the solubility when increasing the molar fraction of alkaline-earth cation as observed at low pressure in \ce{K2CO3}-\ce{CaCO3} mixtures (Figure~\ref{fkcgr}). The observation that the solubility is negatively correlated with the size of the NG still stands at high pressure (at least up to 3~GPa). However beyond 3~GPa, the solubilities of Ne and He come close to each other in natrocarbonatite and even tend to cross at about 6~GPa (He shows a solubility maximum at about 2~GPa), features which are also observed in dolomite. In the same way, the solubility of Xe in natrocarbonatite increases steadily and becomes as high as the one of Ar at 6~GPa, the latter leveling off at about 1~GPa.\\
In the natrocarbonatite melt, the solubility of NGs first increases quasi linearly with pressure, so Henry's law is fulfilled up to a few kbar (Figure \ref{fsol-gr}). Above $\sim 1$~GPa the solubility generally levels off and it even goes through a maximum for He at $P \sim 2$~GPa and for Ar at $P \sim 4$~GPa. As for Xe, its solubility is quasi linear with $P$ (up to $\sim 2$~GPa). In molten dolomite, the solubility of noble gases barely depends on pressure between 3 and 6~GPa. In comparison the solubility of noble gases in silicate melts increases up to $\sim 5$~GPa.  At higher pressures, some experimental studies report a drop of the solubility\cite{Bouhifd2008} and other do not.\cite{Schmidt2002,Niwa2013} In their simulation study \citet{Guillot2012} show a good agreement with the experimental results up to $\sim 5$~GPa and predict a plateau value for the solubility with a slowly decreasing behavior at higher pressures. This trend is similar to what we observe in molten carbonates (Figure~\ref{fsol-gr}), although the solubility plateau occurs at a lower pressure for carbonates. In any case at the pressures of the Earth's upper mantle, our results point out that the solubility of the nobles gases has the same order of magnitude in molten carbonates and in molten silicates \cite{Guillot2012}. This suggests that noble gases would not partition massively in a carbonated phase. However, seeing the significant dependence  of the NG solubilities with the composition of the melt (a factor of $\sim 3$ for Ar in dolomitic versus natrocarbonatitic melts, see the inset of Figure~\ref{fsol-gr}) the investigation of the partition coefficients between carbonate and silicate deserves a further studying. Still, considering that carbonatite melts only represent a minor fraction of magmatic liquids, they can unlikely be the main carrier of noble gases.
\FloatBarrier
\section{Surface Tension}\label{sstens}
\label{sec:stenP}
\begin{figure}
\centering
\includegraphics{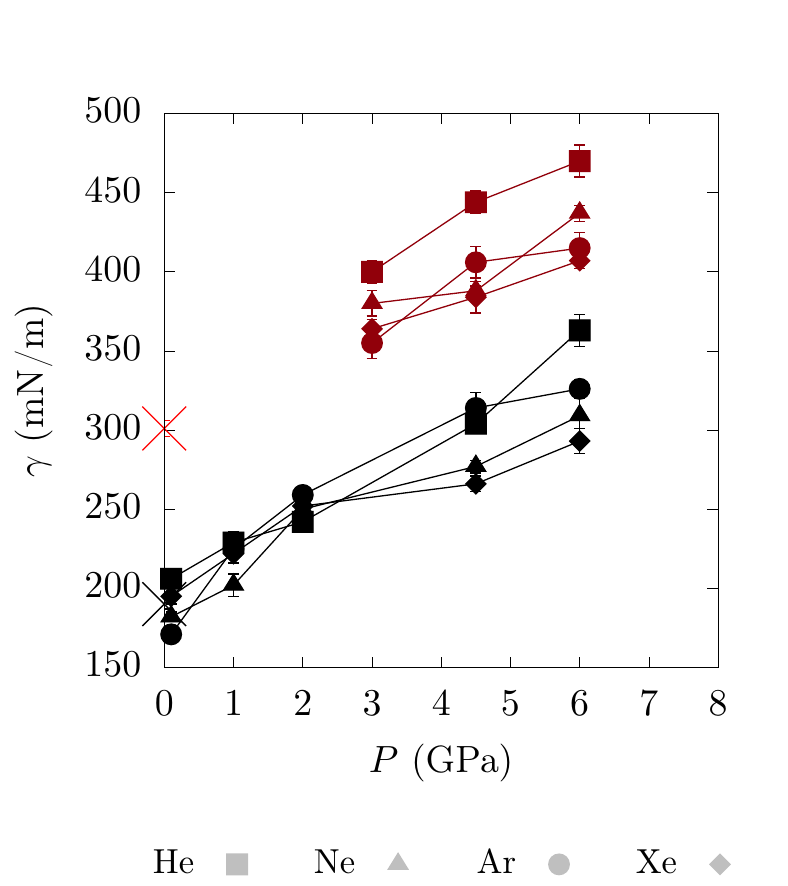}
\caption{Pressure evolution of the surface tension between a noble gas fluid and the natrocarbonatite at 1600~K (black) and molten dolomite at 1673~K (brown). The crosses represent the surface tension of the  natrocarbonatite at 1600~K (black) and of \ce{CaCO3} at 1623~K (red) at the interface with a vacuum (190 $\pm$ 3 mN$\cdot$m$ ^{-1}$ and 301 $\pm$ 5 mN$\cdot$m$ ^{-1}$, respectively).}
\label{fgamma-gr}
\end{figure}
To our knowledge, there are no data of the pressure evolution of the surface tension at the interface between a noble gas phase and molten carbonates, despite the interest in a geochemical perspective. For example if we consider a gas-saturated magmatic melt, it is the liquid/gas surface tension that controls the formation, the growth and the coalescence of gas bubbles below a certain supersaturation pressure \cite{Mangan2005}.  \\ 
By simulating a biphasic system in MD as shown in Figure~\ref{fdprofile} for the calculation of the solubility by the EIM, the surface tension $\gamma$ can be calculated simultaneously for the same computational cost. It is given by the long time limit of the average of the diagonal components of the stress tensor:\cite{Kirkwood1949}
\begin{equation}
\begin{split}
& \gamma =\underset{t \to \infty}{\lim} \gamma(t)\,,\, \\ &\text{with}\,\gamma(t) = \frac{L_z}{2}\Big\langle\Pi_{zz}-\frac{1}{2}\big(\Pi_{xx}+\Pi_{yy}\big)\Big\rangle \enspace , \end{split}
\label{est}
\end{equation}
where $L_z$ is the length of the simulation box perpendicular to the interface and $\Pi_{xx}$, $\Pi_{yy}$ and $\Pi_{zz}$ are the diagonal components of the stress tensor evaluated in the entire simulation box (see~\citet{moi2018} for more details).\\
Figure~\ref{fgamma-gr} plots the surface tension between the natrocarbonatitic (or dolomitic) melt and the noble gas phase (He, Ne, Ar and Xe), calculated at different pressures. For natrocarbonatite at 1600~K the surface tension was also calculated at the interface with a vacuum ($190~\pm~3$~mN/m). This value compares well with the ones obtained at the interface with noble gases at 0.1~GPa ($171-206$~mN/m). \\ 
Irrespective of the nature of the melt and of the noble gas considered, the surface tension increases with increasing pressure. In contrast, it is known that the surface tension of silicate melts in contact with \ce{H2O} decreases with pressure.\cite{Colucci2016} The behavior of carbonate melts may somehow appear as counter-intuitive because with increasing pressure the density of the gas phase approaches that of the melt, thus the energetic cost to create the interface should decrease. However the surface tension involves subtle mechanisms which are likely pressure-dependent. Among other factors, the amount of gas solubilized in the melt which increases with pressure (see previous section) could be responsible for a non-negligible contribution (positive or negative) to the surface stress. On the one hand, it could tend to destabilize the melt (and thus decrease $\gamma$), but on the other hand the gas phase in contact with the melt can act like a piston and rigidify the interface (and thus increase $\gamma$). In any case, the solubilized gas is likely not the only factor in action as the rather large differences of solubility observed between the four gases (almost a factor of ten between He and Xe, see Figure~\ref{fsol-gr}) are not retrieved for the surface tension (see Figure~\ref{fgamma-gr}).\\
There is also a distinct effect of the melt composition on the surface tension: For a given noble gas species and at a given pressure, the surface tension of dolomite is greater than the one of the natrocarbonatite by $\sim$~50\%. This increase of the surface tension with the amount of alkaline-earth cations in the melt is also observed at the interface with a vacuum\cite{moi} (compare the values of $\gamma$ for the natrocarbonatite and for the \ce{CaCO3} melt on Figure~\ref{fgamma-gr}).\\
As for the effect of the gas composition, it is trickier to decipher. At a given pressure and for a given melt composition, there is no systematic trend of the surface tension as a function of the size of the noble gas atom (except for He in dolomite, see Figure~\ref{fgamma-gr}). Moreover, when pressure is increased the hierarchy between noble gases seems to modify somewhat. It is possible that the uncertainties on the calculated values are greater than the ones we estimate from the fluctuations ($\sim 5 \%$, see Figure~\ref{fconvstens} in the appendix) of the running average of Eq.~(\ref{est}).

\section{Conclusion}
To complete and go beyond the precursory study of Burnard \emph{et al.}\cite{Burnard2010}, the solubility of noble gases (He, Ne, Ar and Xe) in carbonate melts was calculated by molecular dynamics simulations. These simulations used empirical interaction potentials whose accuracy was previously demonstrated in studying the thermodynamic and transport properties of carbonate melts and the solubility of noble gases in silicate melt.\cite{moi,moi2018,moi2019,Guillot2012}\\
The NG solubilities were first calculated in \ce{K2CO3}-\ce{CaCO3} mixtures at 1 bar and the results are in a fair agreement with the data of Burnard \emph{et al.}\cite{Burnard2010} once considered the uncertainties on the experimental values. Then we investigated the effect of pressure (up to 6~GPa), focusing on two melt compositions: a dolomitic one and a mixture modeling the carbonatitic lava emitted at Ol Doinyo Lengai. We observed that the solubility decreases with the amount of alkaline-earth cation in the melt and with the size of the noble gas (entropy-driven solubility). Concerning the solubility in the natrocarbonatitic melt, Henry's law is fulfilled at low pressures ($\sim 0.1$~GPa). At higher pressures (a few GPa) the solubility levels off or even starts to diminish smoothly (for He at $P > 2$ GPa and Ar at $P > 4$ GPa). In contrast, in molten dolomite the effect of pressure is negligible on the studied $P$ range ($3-6$~GPa).
At the pressures of the Earth's mantle, the solubilities of noble gases in carbonate melts are still of the same order of magnitude as the ones in molten silicates, a finding in agreement with the ratio \textsuperscript{4}He/\textsuperscript{40}Ar measured in the gas emitted from  Ol Doinyo Lengai crater in 2005, which is close to the mantle value (see \citet{Fischer2009}). Furthermore this finding also suggests that carbonatitic melts at depth cannot be preferential carriers of noble gases.\\
Finally we provided some insight into the surface tension at the interface between carbonate melts and noble gases. With increasing pressure ($P$ from 0 to 6~GPa), the surface tension increases (by a factor $\sim 2$), whatever the composition of the melt and of the NG phase. This is in strong contrast with the effect of \ce{H2O} on the surface tension of silicate melts which drops when pressure increases to a few GPa.
\begin{acknowledgments}
The research leading to these results has received funding from the R\'{e}gion Ile-de-France and the European Community’s Seventh Framework Program (FP7/2007-2013) under Grant agreement (ERC, N$^\circ$ 279790).
\end{acknowledgments}

\appendix

\setcounter{equation}{0}
\setcounter{figure}{0}
\setcounter{table}{0}
\renewcommand{\theequation}{\Alph{section}.\arabic{equation}}
\renewcommand{\thefigure}{\Alph{section}.\arabic{figure}}
\renewcommand{\thetable}{\Alph{section}.\arabic{table}}
\FloatBarrier
\section{Force Field}
\label{a-FF}
The interactions between two noble gas (NG) atoms $i$ and $j$ were modeled by the FF used by \citet{Guillot2012} and consisting of Buckingham potentials: 
\begin{equation}
 V_{ij}(r_{ij})={A_{ij}}\exp(-{r}_{ij}/{\rho_{ij}})-C_{ij}/r_{ij}^6 \enspace .
\label{ebuck}
\end{equation}
Values of the parameters for these interactions are collected in Table~\ref{tffng}. 
\begin{table}[h]
\centering
\begin{tabular}{|l|r|l|r|}
\hline 
NG  & $A_{ij}$ (kJ/mol)  & $\rho_{ij}$ (\AA)  & $C_{ij}$ (\AA$^{6}$/mol) \tabularnewline \hline
He & 132917.0 & 0.2051 & 109.84 \\ \hline
Ne & 684325.4 & 0.2083 & 523.19 \\ \hline
Ar & 2947863.0 & 0.2485 & 5607.64 \\ \hline
Xe & 4474435.5 & 0.2940 & 27142.12 \\ \hline
\end{tabular}
\caption{Buckingham parameters for the NG-NG interactions.\cite{Guillot2012}}
\label{tffng} 
\end{table} 

For the interactions within the carbonate melts we used the FF developed in \citet{moi2018,moi2019}. The intramolecular potential energy associated with a carbonate molecule anion consists of an oxygen-oxygen repulsive potential as in Eq.~(\ref{ebuck}) and of a carbon-oxygen potential (harmonic stretching + coulombic interaction): $
V_{\text{CO}}^{intra}(r_{\text{CO}})=\frac{1}{2}{k_{\text{CO}}}(r_{\text{CO}}-{r_{0,\text{CO}}})^{2}+{q_{\text{O}}q_{\text{C}}}/{4\pi\epsilon_{0}r_{\text{CO}}}$. Moreover, two elements $i$ and $j$, with $i,j$= Na, K, Ca, Mg, O and C (with O and C not belonging to a same carbonate group) interact through a pair potential: $ V_{ij}(r_{ij})={A_{ij}}\exp(-{r}_{ij}/{\rho_{ij}})-C_{ij}/r_{ij}^6+ {q_{i}q_{j}}/{4\pi\epsilon_{0}{r}_{ij}} $, that is a sum of a van der Waals (Buckingham-like) and of a coulombic term. All the parameters for the melt-melt interactions are summarized in Table~\ref{tffsh}.\\
\begin{table}[h]
\centering
\begin{tabular}{|l|l|r|l|r|r|}
\hline 
$i$  & $j$  & $A_{ij}$ (kJ/mol)  & $\rho_{ij}$ (\AA)  & $C_{ij}$ (\AA$^{6}$/mol)  & $q_{i}$ (e) \tabularnewline
\hline 
Mg  & O  & 243 000  & 0.24335  & 1 439  & +1.64202 \tabularnewline \hline 
Ca  & O  & 200 000  & 0.2935  & 5 000  & +1.64202 \tabularnewline \hline 
Na  & O  & 1 100 000  & 0.2228  & 3 000  & +0.82101 \tabularnewline \hline 
K  & O  & 900 000  & 0.2570  & 7 000  & +0.82101 \tabularnewline \hline
C  & O  & 0  & 1  & 0  & +1.04085 \tabularnewline  \hline
O  & O  & 500 000  & 0.252525  & 2 300  & $-$0.89429 \tabularnewline 
\hline 
\end{tabular}\caption{Intermolecular potential parameters and partial charges for the carbonate melts. Intramolecular repulsion parameters between oxygen atoms are
$A_{{\rm {OO}}}^{intra-{\rm {CO}_{3}}}=2611707.2$~kJ/mol and $\rho_{{\rm {OO}}}^{intra-{\rm {CO}_{3}}}=0.22$~\AA.\cite{moi2018,moi2019}}
\label{tffsh} 
\end{table}

For the NG-melt interactions we used Lennard-Jones potentials: 
\begin{equation}
 V_{ij}(r_{ij})=4{\epsilon_{ij}} \Big( \big({\sigma_{ij}}/{{r}_{ij}}\big)^{12}-\big({\sigma_{ij}}/{{r}_{ij}}\big)^6\Big) \enspace ,
 \label{elj}
\end{equation} 
as adjusted by \citet{Guillot2012} and by Aubry \emph{et al.}\cite{Aubry2013} (Table~\ref{tffngm}).
\begin{table}[hbt]
\centering
\begin{tabular}{|l|r|r|r|r|r|r|r|r|r|}
\hline 
  & $\epsilon_{\rm{He}}$  & $\sigma_{\rm{He}}$
  & $\epsilon_{\rm{Ne}}$   & $\sigma_{\rm{Ne}}$ 
  & $\epsilon_{\rm{Ar}}$   & $\sigma_{\rm{Ar}}$ 
  & $\epsilon_{\rm{Xe}}$   & $\sigma_{\rm{Xe}}$    \tabularnewline
\hline 
Mg &1.554 &2.076 &2.896 &2.125 &3.897 &2.455  &4.068 &2.746 \\ \hline
Ca &1.142 &2.465 &2.142 &2.513 & 3.389 &2.843 & 3.943 &3.134  \\ \hline
Na &0.523 &2.552 &0.984 &2.600 &1.596 &2.930  &1.891 &3.221  \\ \hline
K &0.351 &3.008 &0.663 & 3.056 &1.232 & 3.386  &1.603&  3.677 \\\hline
O & 0.525 & 2.81 & 0.671 & 2.89 & 1.131 & 3.19 & 1.317 & 3.44 \\ \hline
\end{tabular}\caption{Lennard-Jones parameters for noble gas - carbonate melt interactions with $\epsilon$ in kJ/mol and $\sigma$ in \AA.\cite{Guillot2012,Aubry2013}}
\label{tffngm} 
\end{table}
\FloatBarrier
\section{The Test Particle Method}
\setcounter{figure}{0}
\setcounter{table}{0}
\begin{figure}[htb]
\centering
\includegraphics[scale=0.85]{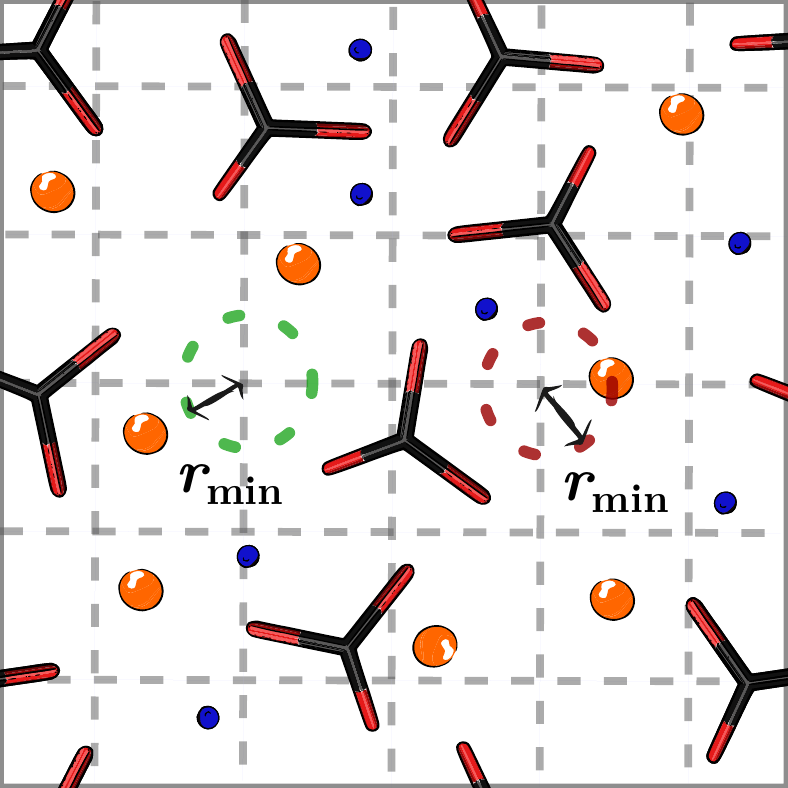}
\caption{Schematic representation of the test-particle method on given 2D configuration of the solvent. A meshing of the simulation box is defined (e.g. $L/$d$L=5$, gray dashed line) then the insertion of the test-particle is attempted on each node (dashed circles). The interaction energy between the solute and the solvent $\Psi_i$ is calculated under the condition that no solvent atom is located within a previously defined cutoff radius $r_{min}$. If that radius overlaps a solvent atom (red circle), $\Psi_i$ is very large (repulsive configuration) and its contribution to $\text{e}^{-\Psi_i/k_BT}$ is null. Only the events where the test-particle is inserted within a solvent cage are significant for the solubilty (green circle).}
\label{fwidom}
\end{figure}
\subsection{Thermodynamics}
\label{a-thermo}
Calculating the solubility of a noble gas within a melt consists in considering a thermodynamic equilibrium between the gas and the melt (in contact with each other), expressed by the equality of the chemical potentials of the gas in both phases: 
\begin{equation}
\mu_v=\mu_m \enspace .
\label{eeqthermo}
\end{equation}
The chemical potentials can be decomposed as the sum $\mu_{v,m}=\mu_{v,m}^{IG}+\mu_{v,m}^{ex}$, of an ideal gas part 
\begin{equation}
\mu_{v,m}^{IG}= C + k_BT\ln \rho_{v,m}^g \enspace ,
\label{eb2}
\end{equation}
where $\rho^g$ is the number density of the noble gas phase in its pure phase ($\rho_{v}^g$) or in the melt ($\rho_{m}^g$) and $C$ is a constant that is equal in both phases (gas and melt), and an excess part $\mu^{ex}$ ($\mu_{v}^{ex}$ in the gas phase, $\mu_{m}^{ex}$ in the melt). Introducing Eq.~(\ref{eb2}) into Eq.~(\ref{eeqthermo}) leads to: 
\begin{equation}\label{aegammatx}
\frac{\rho^g_m}{\rho^g_v}=\mathrm{e}^{-\left(\mu_m^{ex} - \mu_v^{ex}\right)/k_BT}=\frac{\Gamma_m}{\Gamma_v} \enspace ,
\end{equation}
where $\rho_m^g$ and $\rho_v^g$ are the number densities of the gas in the melt and in the vapor phase, respectively, and  $\Gamma_m$ and $\Gamma_v$ are the solubility parameters of the gas in the two phases.\\
The solubility of the gas in the melt, expressed as the number of moles of gas in the melt divided by the number of moles of melt and of noble gas, is given by
\begin{equation}
X=\frac{\frac{\rho_v}{\rho_m} \cdot \frac{\Gamma_m}{\Gamma_v}}{1+\frac{\rho_v}{\rho_m} \cdot \frac{\Gamma_m}{\Gamma_v}} \enspace ,
\label{aegammax}
\end{equation}
where $\rho_v$ is the number density of the gas phase and $\rho_m$ is the number density of the melt.\\
At low pressure, the gas phase can be considered as ideal: $\mu_v^{ex}=0$ (i.e. $\Gamma_v=1$ and $P~=~\rho_vk_BT$) thus $\rho_m^{g}=\Gamma_m\rho_v^{g}=\Gamma_mP/k_BT$.  Then from equation (\ref{aegammax}) it follows: 
\begin{equation}
X=P\Gamma_m/\rho_mk_BT \enspace ,
\label{aehenry}
\end{equation}
which is nothing but Henry's law. In this case, only the solubility parameter of the noble gas in the melt, $\Gamma_m$, has to be calculated.

\FloatBarrier
\subsection{Simulation details}
\label{a-TPM}
\begin{figure}[hbt]
\centering
\includegraphics{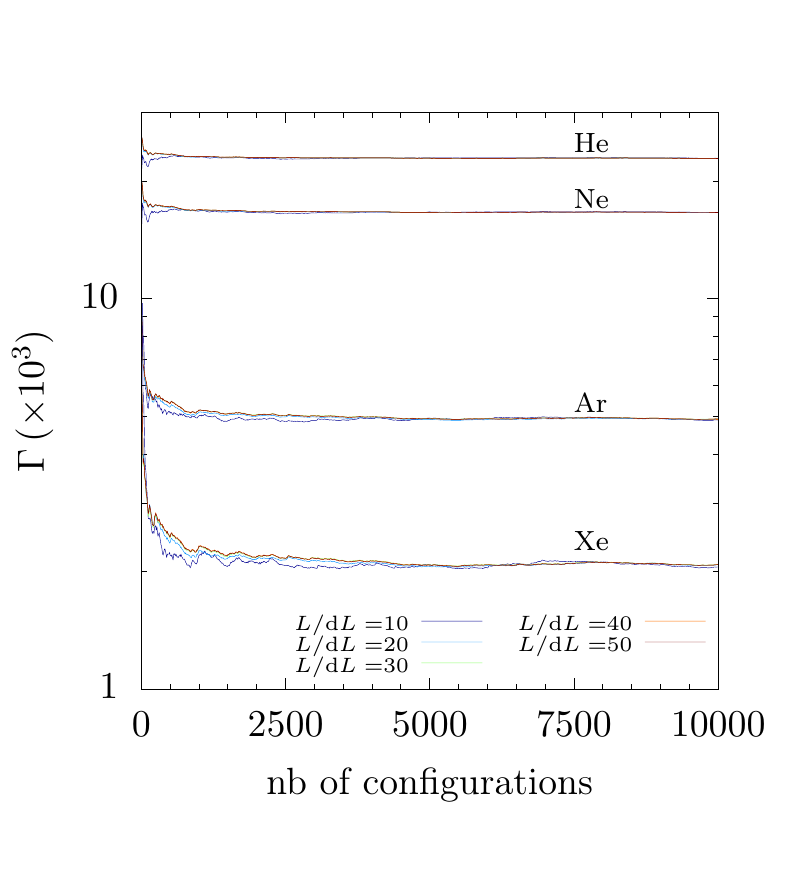}
\caption{Solubility parameter $\Gamma = \langle \text{e}^{-\Psi/k_BT}\rangle$   of He, Ne, Ar and Xe in  a carbonate melt (e.g. natrocarbonatite at 1600~K and 0.1~GPa) calculated as a cumulative average over simulation length. The distance between two attempted insertions of the particle is given by d$L$, $L$ is the length of the simulation box (see text).}
\label{fconvtpm}
\end{figure}

Classical MD simulations of the carbonate melt ($N\simeq$ 1000 atoms) were carried out using the DL\_POLY~2 software \cite{Smith1996}, with a timestep of 1~fs. The simulations were performed in the $NVE$ ensemble with an equilibration run of 0.5~ns, followed by a production run of at least 10~ns. Insertion of the test particle in the simulation box was attempted every 1000 timesteps (1 ps) on a grid of width $\rm{d}L$, with $L/\rm{d}L=10$ to 50 (i.e. the grid has $10^3=1000$ to $50^3 = 125000$ nodes, Figure~\ref{fconvtpm} evidences that convergence is reached for $L/\rm{d}L \ge 30$). In practice, if the distance between the node and a solvent atom is too small (within a defined cutoff radius $r_{min}$, see Figure~\ref{fwidom}), the insertion is rejected and a null contribution is added to the average of Eq.~(\ref{egammanvt}) (because the value of $\mathrm{e}^{-\Psi /k_BT}$ in Eq.~(\ref{egammanvt}) is vanishingly small for this event). Note that when the density of the solvent is high or when the inserted particle has a large van der Waals radius, the rate of rejection for the insertion is high and uncertainties may become important(e.g. Ne, Ar and Xe in \ce{K2CO3}, see Figure~\ref{fkcgr}). In this study, the solubility calculations using this method have uncertainties of a few percents. For more detailed applications with MD, see \citet{Guillot2012} and Aubry \emph{et al.}\cite{Aubry2013}.
\FloatBarrier
\subsection{Data}
\FloatBarrier
\begingroup
\squeezetable
\begin{table}[ht]
\tiny
\begin{tabular}{l l r r r r} 
		& $\rho_m$ &	$\Gamma_m^{\rm{He}}$ & $\Gamma_m^{\rm{Ne}}$	& $\Gamma_m^{\rm{Ar}}$	& $\Gamma_m^{\rm{Xe}}$\\ 
		&(mol/L) 	& ($\times 10^{3}$)	& ($\times 10^{3}$)	&  ($\times 10^{3}$)	& ($\times 10^{3}$) \\ \hline
\\
KC \\
$x_{\rm{K}_2\rm{CO}_3}= 0$ & 24.08 & $2.69 \pm 0.03$ & $1.74 \pm 0.04$ & $0.27 \pm 0.01$ & $0.059 \pm 0.05$
\\
$x_{\rm{K}_2\rm{CO}_3}= 0.25$ & 20.12 & $6.70 \pm 0.02$ & $4.43 \pm 0.02$ & $0.96 \pm 0.02$ & $0.26 \pm 0.02$
\\
$x_{\rm{K}_2\rm{CO}_3}= 0.33$ & 19.39 & $6.65 \pm 0.03$ & $4.24 \pm 0.03$ & $0.80 \pm 0.02$ & $0.17 \pm 0.03$
\\
$x_{\rm{K}_2\rm{CO}_3}= 0.4$ & 19.67 & $7.29 \pm 0.05$ & $4.60 \pm 0.04$ & $0.83 \pm 0.02$ & $0.17 \pm 0.02$
\\
$x_{\rm{K}_2\rm{CO}_3}= 0.5$ & 17.66 & $8.63 \pm 0.03$ & $5.47 \pm 0.03$ & $1.08 \pm 0.03$ & $0.26 \pm 0.03$
\\
$x_{\rm{K}_2\rm{CO}_3}= 1$ & 13.77 & $16.00 \pm 0.07$ & $10.62 \pm 0.06$ & $2.66 \pm 0.07$ & $ 0.79 \pm 0.03$
\\
\\ \hline \\
NKC & \\
$P = 1$~bar & 17.64 & $27.97 \pm 0.05$ & $20.90 \pm 0.05$ & $6.95 \pm 0.06$ & $3.22 \pm 0.06$ \\
$P = 0.1$~GPa & 17.98 & $22.91 \pm 0.04$ & $16.66 \pm 0.04$ & $4.94 \pm 0.03$ & $2.09 \pm 0.03$ \\
\\ \hline 
\end{tabular}
\caption{Solubility parameters of NGs in melts: \ce{K2CO3}--\ce{CaCO3} mixtures with different molar ratios of \ce{K2CO3} at 1173~K and 1~bar and natrocarbonatite (Na$_{1.1}$K$_{0.18}$Ca$_{0.36}$CO$_3$, NKC) at 1600~K, calculated by MD using the TPM.}
\label{tanntpm}
\end{table}
\endgroup
\FloatBarrier
\section{The Explicit Interface Method}
\setcounter{figure}{0}
\setcounter{table}{0}
\subsection{Simulation details}
\label{a-EIM}
\begin{figure}[hbt]
\centering
\includegraphics{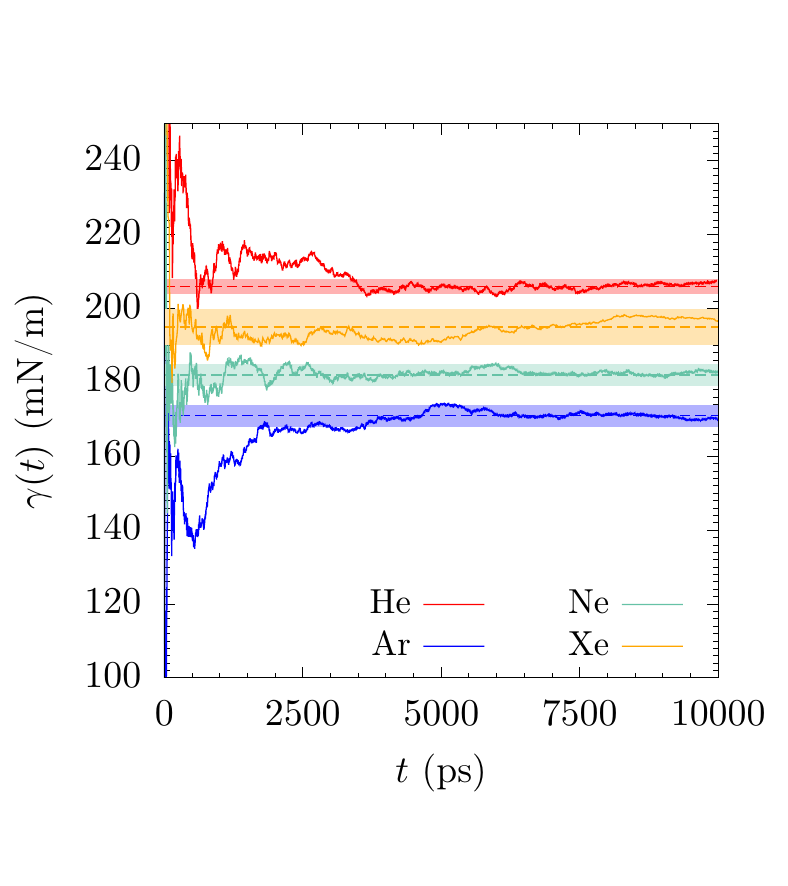}
\caption{The surface tension at the interface between a noble gas phase (He, Ne, Ar and Xe) and a carbonate melt (e.g. natrocarbonatite at 1600~K and 0.1~GPa) is calculated from the cumulative average over simulation length $\gamma(t) = \frac{L_z}{2}\Big\langle\Pi_{zz}-\frac{1}{2}\big(\Pi_{xx}+\Pi_{yy}\big)\Big\rangle$ (see Eq.~(\ref{est}) of main text). The dashed lines represent the converged values of the surface tension $\gamma$ within error bars (shaded areas) estimated from the fluctuations of $\gamma(t)$.}
\label{fconvstens}
\end{figure}

The EIM simulations consisted in modeling the solvent ($N\simeq$ 2000 atoms of melt) in contact with a NG phase in a parallelepipedic simulation box. The number of NG atoms in the box ($N\simeq$ 300 - 900 atoms) was chosen so that the vapor phase was large enough for the consecutive periodic images of the melt to not interact with one another. \\
An equilibration run was first performed in the $NPT$ ensemble (using a Nos\'e-Hoover thermostat) for 0.9~ns (including 0.5~ns to equilibrate the temperature) allowing to reach an accuracy on the density value of $\Delta n/n \sim \pm1\%$ for the two coexisting phases. 
To evaluate the solubility $X$, simulations were performed in the $NVE$ ensemble with an equilibration run of 0.5~ns, followed by a production run of 10~ns. Configurations were extracted every 1~ps to determine the density profiles and calculate the surface tension between the NG fluids and the carbonate melts (Figure~\ref{fconvstens}). 
\FloatBarrier
\subsection{Data}
\begingroup
\squeezetable
\begin{table}[hbt]
\begin{tabular}{l l l l l l} 
		& $P$ &	$X\textsubscript{He}$ & $X\textsubscript{Ne}$	& $X\textsubscript{Ar}$	& $X\textsubscript{Xe}$\\ 
		&(GPa) 	& (mol\%) 	& (mol\%) 	& (mol\%) 	& (mol\%)\\ \hline
\\
NKC & 
0.1    & 1.5 $\pm$ 0.05  & 0.84 $\pm$ 0.1 & 0.28 $\pm$ 0.1 & 0.16 $\pm$ 0.1 \\
&1.0  &  7.5 $\pm$ 0.5 & 5.2 $\pm$ 0.20 & 1.7 $\pm$ 0.1 & 0.5 $\pm$ 0.15 \\
&2.0 & 9.0 $\pm$ 0.2 & 6.6 $\pm$ 0.4 & 1.8 $\pm$ 0.10  & 1.0 $\pm$ 0.15 \\
&4.5 & 8.3 $\pm$ 0.2 & 7.3 $\pm$ 0.3 & 1.95 $\pm$ 0.15 & 1.4 $\pm$ 0.15\\
&6.0 & 7.5 $\pm$ 0.5 & 7.4 $\pm$ 0.3 & 1.35 $\pm$ 0.1 & 1.6 $\pm$ 0.15\\
\\ \hline \\
CM	& 3.0   & 3.1 $\pm$ 0.4 & 2.7 $\pm$ 0.4 & 0.45 $\pm$ 0.05            & 0.12 $\pm$ 0.03 \\
& 4.5 &  3.1 $\pm$ 0.4 & 3.0 $\pm$  0.3 & 0.55 $\pm$ 0.05 & 0.17 $\pm$ 0.06 \\
& 6.0    &  2.7 $\pm$  0.2 & 2.9 $\pm$ 0.4 & 0.55 $\pm$ 0.1 & 0.15 $\pm$ 0.05 \\
\\ \hline 
\end{tabular}
\caption{Solubility of NGs in melts: natrocarbonatite (Na$_{1.1}$K$_{0.18}$Ca$_{0.36}$CO$_3$, NKC) at 1600~K and dolomite (Ca$_{0.5}$Mg$_{0.5}$CO$_3$, CM) at 1623~K, calculated using the EIM.}
\label{tannX}
\end{table}
\endgroup

\begingroup
\squeezetable
\begin{table}[hbt]
\begin{tabular}{l l l l l l} 
		& $P$ &	$\gamma\textsubscript{He}$ & $\gamma\textsubscript{Ne}$	& $\gamma\textsubscript{Ar}$	& $\gamma\textsubscript{Xe}$\\ 
		&(GPa) 	& (mN/m) 	& (mN/m) 	& (mN/m) 	& (mN/m)\\ \hline
\\
NKC & 
0.1    &
206 $\pm$  2   & 182 $\pm$ 3  &171 $\pm$ 3 & 195 $\pm$ 5 \\
&1.0  &   229  $\pm$ 7  & 202 $\pm$ 7 & 225 $\pm$ 6  & 222 $\pm$ 6 \\
&2.0 & 242 $\pm$ 4  & 250 $\pm$ 8  & 259 $\pm$ 4 & 252 $\pm$ 5\\
&4.5 &    304 $\pm$ 4  & 277 $\pm$ 4  & 314 $\pm$ 10  & 266  $\pm$ 5 \\
&6.0  & 363 $\pm$ 10 & 309 $\pm$ 15 & 326 $\pm$ 6 & 293 $\pm$ 8 \\
\\ \hline \\
CM	& 3.0    &
398  $\pm$ 7  &     
381  $\pm$ 9  & 
357  $\pm$ 6 &
364  $\pm$ 8 \\
& 4.5   &  
444  $\pm$ 9  &
388  $\pm$ 7  &
406 $\pm$ 12  &
382 $\pm$ 11 \\
& 6.0    &   
468 $\pm$ 10  &
435 $\pm$ 9   &
413 $\pm$ 11  &
407 $\pm$ 10 \\
\\ \hline 
\end{tabular}
\caption{Surface tension between NG fluids and melts: natrocarbonatite (Na$_{1.1}$K$_{0.18}$Ca$_{0.36}$CO$_3$, NKC) at 1600~K and dolomite (Ca$_{0.5}$Mg$_{0.5}$CO$_3$, CM) at 1623~K.}
\label{tannsten}
\end{table}
\endgroup

\FloatBarrier
\bibliography{main}
\end{document}